\documentclass[twocolumn, 10pt]{article}  

\usepackage{amssymb}          
\usepackage{graphicx}         
\usepackage[latin1]{inputenc} 
\usepackage{amsmath}          

\usepackage{subfigure}
\usepackage{amsfonts}

\usepackage{latexsym}
\usepackage{graphicx}
\usepackage{color}

\newtheorem{theorem}{Theorem}[section]

\newtheorem{lemma}{Lemma}[section]

\newcommand{\qed}{\hfill $\Box$ \vskip 2ex}

\usepackage[figuresright]{rotating}
\usepackage{enumerate}

\newcommand{\ket}[1]{|#1\rangle}
\newcommand{\bra}[1]{\langle#1|}

\newcommand{\Hi}{\mathcal{H}}
\newcommand{\Si}{\mathcal{S}}

\newcommand{\C}{\mathbb{C}}
\newcommand{\CNOT}{C}

\newcommand{\Span}{\textrm{span}}

\newcommand{\trace}{\textrm{tr}}

\newcommand{\herm}{\mathfrak{H}}
\newcommand{\dens}{\mathfrak{D}}

\newcommand{\beq}{\begin{equation}}
\newcommand{\eeq}{\end{equation}}
\newcommand{\beqa}{\begin{eqnarray}}
\newcommand{\eeqa}{\end{eqnarray}}
\newcommand{\beqan}{\begin{eqnarray*}}
\newcommand{\eeqan}{\end{eqnarray*}}

\title{\LARGE \bf Quantum State Preparation by Controlled Dissipation in Finite Time:\\
From Classical to Quantum Controllers 
\thanks{Partially supported by the QUINTET and QFUTURE strategic research projects of the 
Department of Information Engineering and University of Padua, Italy.}}
\author{
Giacomo Baggio\thanks{G. Baggio is with the Dipartimento di Tecnica e Gestione dei 
Sistemi Industriali, Universit\`a di Padova, Stradella S. Nicola 3, Vicenza, Italy 
({\tt giacomo.baggio@studenti.unipd.it}). },{ }%
Francesco Ticozzi\thanks{F. Ticozzi is with the Dipartimento di Ingegneria
dell'Informazione, Universit\`a di Padova, Via Gradenigo 6/B,
35131 Padova, Italy, and with the 
Department of Physics and Astronomy, Dartmouth College, 6127 Wilder Laboratory, 
Hanover, NH 03755, USA ({\tt ticozzi@dei.unipd.it}).}~%
{ }and Lorenza Viola\thanks{L. Viola is with the Department of Physics and Astronomy, 
Dartmouth College, 6127 Wilder Laboratory, Hanover, NH 03755, USA 
({\tt lorenza.viola@dartmouth.edu}).}}

\date{\today}

\begin{document}
\maketitle

\begin{abstract}
We propose a general scheme for dissipatively preparing arbitrary pure quantum states  
on a multipartite qubit register in a finite number of basic control blocks. 
Our ``splitting-subspace'' approach relies on control resources that are 
available in a number of scalable quantum technologies 
(complete unitary control on the target system, an ancillary resettable qubit 
and controlled-not gates between the target and the ancilla), and can be seen 
as a ``quantum-controller'' 
implementation of a sequence of classical feedback loops.  We
show how a large degree of flexibility exists in engineering the required 
conditional operations, and make explicit contact with a stabilization 
protocol used for dissipative quantum state preparation and entanglement 
generation in recent experiments with trapped ions.    
\end{abstract}

\vspace*{2mm}

\section{Introduction and Notation}

State preparation problems are of vital importance across quantum information 
processing applications, ranging from initialization of quantum computation and 
simulation algorithms in network-based as well as cluster-state architectures to 
use within enhanced quantum metrology protocols \cite{nielsen-chuang}. 
If the preparation is to be achieved irrespective of the initial state of the 
system, from a control standpoint this translates naturally into a stabilization 
problem. Physically, one is compelled to make the quantum evolution irreversible, 
by introducing open-system features in either an open-loop or closed-loop fashion 
\cite{altafini-tutorial}. 
Closed-loop control based on quantum measurements and feedback techniques 
provides, in particular, a very natural and powerful toolbox \cite{wiseman-book}, 
with feedback from a single ancilla qubit together with fast, complete unitary control 
allowing in principle to engineer arbitrary open-system dynamics \cite{viola-engineering}.

Quantum state stabilization problems have been studied in depth for continuous-time 
dynamical models, for either Markovian ``output feedback'' 
\cite{wang-wiseman,ticozzi-markovian,Ticozzi-QL} or strategies based on state 
reconstruction by quantum filtering \cite{vanhandel-feedback,mirrahimi-stabilization}. 
It is worth remarking that in such continuous-time scenarios the target state can 
typically be reached only asymptotically, in the (formal) limit of infinite evolution 
time.  Recently, a linear-algebraic framework for analysis and synthesis has also been 
developed for the discrete-time case.  In particular, for a given indirect quantum 
measurement and 
complete unitary control over the target system, it has been shown that pure states 
are generically stabilizable \cite{bolognani-arxiv}.  This is always true for a 
projective measurement, and if the latter is associated to a non-degenerate observable, 
it further follows that the desired state can be prepared in a {\em single} step of 
measurement-plus-control \cite{bolognani-cyprus}.

From an implementation point of view, a problem associated with measurement-based 
feedback schemes is that the required control resources need not be readily 
available for many state-of-the-art experimental quantum devices.  In this work, 
we investigate how to prepare a desired pure state in {\em finite time} by means of 
a reasonable set of resources for a multipartite qubit system. In particular, 
while still assuming complete unitary control, 
we shall effectively ``encode'' the whole feedback loop in a finite sequence of 
{\em coherent feedback} actions \cite{lloyd-coherent, wiseman-alloptical}, 
where the controller is itself a quantum system (a qubit) and no measurement is 
involved.  For certain experimental settings, most notably trapped ions, these control 
resources are not only achievable in principle, but have been already experimentally 
demonstrated up to 5 qubits \cite{barreiro}.  In particular, we will illustrate how 
the ``stabilizer pumping'' strategy proposed in \cite{barreiro,muller} fits into 
our general ``splitting-subspace'' framework, and how control actions achieving 
stabilization of arbitrary pure target states can be explicitly synthesized. 

We begin by recalling some basic concepts and notations. 
Let $S$ be a {\em finite-dimensional} quantum system of interest, with 
associated Hilbert space $\Hi\sim \C^d.$ Vectors and linear functionals on $\Hi$ are denoted using Dirac's notaion with $\ket{\psi}$ and $\bra{\phi},$ respectively \cite{sakurai}. Observable quantities on $S$ are associated 
to self-adjoint operators on $\Hi$, here represented by Hermitian matrices 
$X=X^\dag\in\herm(\Hi)$.  
In particular, the {\em state} of $S$ is in general 
described by a {\em density operator} 
$\rho\in\dens(\Hi)=\{\rho\in\herm(\Hi)|\rho\geq 0,\;\trace(\rho)=1\}$, 
with {\em pure states} corresponding to the extreme point of the (convex) set 
$\dens(\Hi)$. Unitary matrices are denoted by $U\in{\mathcal U}(\mathcal H).$ 
The (real) spectrum of an observable $X$ represents the set of the possible outcomes 
in the simplest case of a so-called projective (or von Neumann's) quantum measurement 
on $S$. Suppose that $X$ admits a spectral decomposition of the form 
$X=\sum_{i} x_i \Pi_i$, in terms of a complete set of orthogonal projectors 
$\{ \Pi_i\}$ on $\Hi$. 
According to the basic postulates for von Neumann's 
measurements, the probability of obtaining $x_i$ given the pre-measurement 
state $\rho$ is $p_i=\trace(\Pi_i \rho)=\trace(\Pi_i \rho \Pi_i)$. 
Conditionally upon the measurement outcome $x_i$ being recorded, 
the (normalized) post-measurement state of $S$ then becomes 
$\rho|_i=\frac{1}{p_i}\Pi_i\rho\Pi_i.$

If $S$ consists of multiple (distinguishable) {\em subsystems} $S_j$, $j=1,\ldots, N$, 
each associated to a Hilbert space ${\cal H}_j$, the corresponding mathematical 
description is carried out in the tensor product space, $\Hi =\Hi_1\otimes\Hi_2 \ldots 
\otimes \Hi_N$ \cite{nielsen-chuang}, and observables and density operators remain 
associated with Hermitian and positive-semidefinite, trace-one operators on
$\Hi$, respectively.  Given a state $\rho \in \dens(\Hi)$ of $S$, the 
{\em reduced state} of one (or a subset) of subsystems may be uniquely 
determined by taking the partial trace over the 
remaining subsystem(s), for instance in the simplest bipartite setting, $\Hi =\Hi_1 
\otimes \Hi_2$, we shall denote by $\rho_1 = \trace_{{\Hi}_2} (\rho)$ the reduced 
density operator describing the first subsystem alone.  

If information about a quantum system is gathered indirectly, through measurements of a 
correlated auxiliary system, the formalism of von Neumann's projective 
measurements is overly restrictive in general and a description of the 
effect of the measurement on the system of interest is provided by so-called 
{\em generalized measurements}.  To a set of possible measurement outcomes labeled 
by $k$, we associate a set of measurement operators $\{M_k\}$ on the system 
of interest, in such a way that 
$\sum_k M_k^\dag M_k=I,$ with $I$ being the identity operator.
The probability of obtaining the $k$-th outcome is then computed as 
$p_k=\trace(M_k^\dag  M_k\rho)$, $\rho$ denoting now the reduced state of 
the system of interest which, after the outcome is recorded, is updated 
to $\rho|_k=\frac{1}{p_k}M_k\rho M_k^\dag.$

By taking the average over the possible outcomes, we obtain a Trace-Preserving 
Completely Positive (TPCP) linear transformation of the state in the form of a 
so-called Kraus map \cite{kraus}, that is, 
\[ \rho \mapsto {\cal E}(\rho)=\sum_k M_k\rho M_k^\dag.\]
The standard case of projective measurements on $S$ is formally recovered by 
choosing $M_k=\Pi_k.$

\vspace*{2mm}

\section{Discrete-Time Feedback Stabilization}

Suppose that a generalized measurement operation can be performed on the 
target system at times $t=1,2,\ldots$, resulting in an open-system, discrete-time 
dynamics described by a Kraus map with operators $\{M_k\}$. Suppose, in addition, 
that we are able to enact arbitrary unitary control actions on the state of $S$, 
that is, $\rho \mapsto U\rho U^\dag,$ for arbitrary $U\in{\mathcal U}(\mathcal H)$, 
with unitary manipulations that are fast with respect to the measurement 
time scale, so that the measurement and the control effectively act in 
distinct ``time slots''. 

We can then in principle implement a Markovian feedback law, consisting in a map 
from the set of measurement outcomes to the set of unitary
matrices, 
$$U(k):k\mapsto U_k\in{\mathcal U}(\mathcal \Hi).$$ 
If the above measurement-control loop is iterated and we average over the measurement 
results at each step, the net result is a different TPCP map, which describes the 
evolution of the state {\em immediately after} each application of the controls:
$$ \rho(t+1)=\sum_kU_kM_k\rho(t)M_k^\dag U_k^\dag. $$
Controllability and stabilizability for the resulting class of discrete-time, 
closed-loop dynamics have been studied in detail in 
\cite{viola-engineering, bolognani-cyprus,bolognani-arxiv,albertini-controllability}.
In particular, from the results of \cite{bolognani-arxiv,albertini-controllability}, 
it is immediate to see that if the following control resources are available:

\begin{itemize}
\item[(f1)] {\em Arbitrary unitary control actions $\{U_k\}$}; 

\item[(f2)] {\em A non-degenerate projective measurement}, associated to a 
resolution of the identity on $S$, $\{\Pi_k=\ket{\phi_k}\bra{\phi_k}\}_{k=1}^{d},$ 
\end{itemize}

\noindent 
then the system can be prepared in {\em any desired pure state in one step}. 
Let $\rho_d=\ket{\psi}\bra{\psi}$ denote the target pure state: the desired 
preparation is then simply achieved by choosing control operations $U_k$ such that
$ U_k\ket{\phi_k} = \ket{\psi}.$
In control-theoretic terms, this effectively implements a quantum {\em dead-beat} 
controller, reaching the desired state not just asymptotically but in one step.
The above is indeed an abstract description of the most straightforward 
procedure to prepare a given state: first, measure the system projecting it onto some 
known state, and then enact some open-loop, controlled transition to steer it to the 
desired state.  Despite its conceptual simplicity, this strategy may become challenging 
in practical control scenarios where the required measurement procedures are unavailable 
(for instance, measurement may be destructive, too slow and/or inaccurate, or not having 
the needed resolution).  In what follows, by focusing on multi-qubit systems, we show 
how to achieve the same pure-state preparation by replacing the above full resolution 
measurement with the ability of using an extra qubit as a fully coherent resettable 
quantum controller \cite{lloyd-coherent,lloyd-uqi}.

\vspace*{2mm}

\section{The Splitting-Subspace Approach}

Consider an $N$-qubit register, with associated Hilbert space 
$\Hi_Q =\bigotimes_j\Hi_j\sim \C^{2^N},$ and with 
$\{\ket{\phi_j}\}_{j=1}^{2^N}$ denoting the standard (computational) 
basis of $\Hi_Q,$ $\ket{\phi_1}=\ket{0\ldots 0 0},\,\ket{\phi_2}=\ket{0\ldots 0 1}, 
\ket{\phi_3}=\ket{0\ldots 1 0},$ and so on.

Assume that the following control resources are available:

\begin{itemize}
\item[(s1)] {\em Arbitrary unitary control actions $\{U\}$} on 
the $N$ target qubits;

\item[(s2)] {\em An auxiliary control qubit}, with Hilbert space $\Hi_c,$ 
that can be reset to a known pure state, say $\ket{1};$

\item[(s3)] {\em Controlled-not unitaries} $ \CNOT_{\text{in}}, 
\CNOT_{\text{out}} \in {\cal U}(\Hi_c\otimes\Hi_j)$ {\em between the 
control and one of the target qubit}. Without loss of generality, we can 
pick the first qubit ($j=1$) and write  
$\CNOT_{\text{out}}=I_2\otimes\ket{0}\bra{0}+\sigma_x\otimes\ket{1}\bra{1}$, 
$\CNOT_{\text{in}}=\ket{1}\bra{1}\otimes I_2+\ket{0}\bra{0}\otimes\sigma_x$, 
respectively.  
\end{itemize}

We next show that a simple approach can be developed in order to design quantum 
circuits able to prepare the target pure state $\rho_d=|\psi\rangle\langle\psi|$ 
in a finite number of iterations. The first step is to provide a characterization 
of the target state in terms of a family of {\em splitting subspaces}.

\begin{lemma}{\bf (Splitting Subspaces)}
\label{splitsubs} 
Any $N$-qubit pure state $\ket{\psi}\in\Hi_Q$ can be described as the unique 
unit-norm vector in the intersection of $N$ subspaces ${\cal S}_k$ of 
dimension $2^{N-1},$ that is, 
\begin{equation}
\Span\{\ket{\psi}\}=\bigcap_{k=1}^N {\cal S}_k.
\label{SS}
\end{equation}
\end{lemma}

\begin{figure}[t]
\centering
\input{blockdiagram}
\includegraphics[width=8cm]{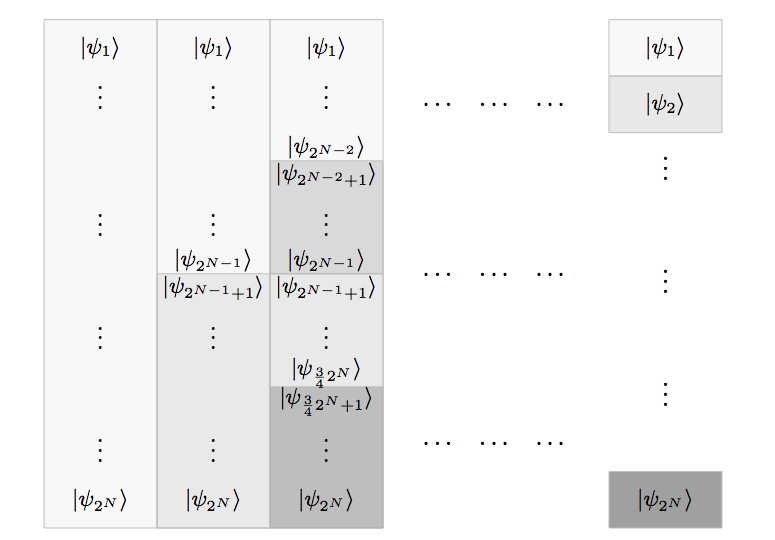} 
\caption{Pictorial representation of the constructive argument used in Lemma 3.1 to 
obtain a splitting-subspace description for the target state $|\psi_1\rangle = 
|\psi\rangle$.}
\label{splitting}
\end{figure}

{\em Proof.} It suffices to provide an explicit way to construct the ${\cal S}_k$, 
see also Figure \ref{splitting} for illustration.  
Start by relabeling $|\psi_1\rangle:=\ket{\psi}$ and complete it with $2^N -1$ vectors 
so that $\{\ket{\psi_k} \}_{k=1}^{2^N}$ is an orthonormal basis\footnote{Clearly, there 
is a lot of freedom in doing so, effectively equivalent to choosing an element of 
$SU(2^N-1)$. This freedom can be exploited in reducing the complexity 
of the unitary operations in what follows.}. 
Next, define 
${\cal S}_1=\Span\{\ket{\psi_k},\;k=1,\ldots,2^{N-1}\}$, 
${\cal S}_2=\Span\{\ket{\psi_k},\;k=1,\ldots,2^{N-2},2^{N-1}+1,\ldots,2^{N-1}+2^{N-2}\}$, 
and so on for the remaining subspaces. Formally, by defining the matrices 
\begin{eqnarray*}
I_2&\hspace*{-1.5mm}=\hspace*{-1.5mm}&\left[\begin{array}{cc}
1 & 0 \\0 & 1\end{array}\right] , \\
\Pi^{(k)}&\hspace*{-1.5mm}=\hspace*{-1.5mm}&I_2^{\otimes (k-1)}\otimes\left[\begin
{array}{cc}1 & 0 \\0 & 0\end{array}\right]\otimes I_2^{\otimes (n-k)},
\end{eqnarray*}
in the $\ket{\psi_k}$-basis, we can identify the $k$-th splitting subspace as 
\[{\cal S}_k=\textrm{range}(\Pi^{(k)}).\]
By construction, the only vector in the intersection of all such subspaces 
is $\ket{\psi_1}.$ 
\qed

For the important class of {\em stabilizer states} \cite{nielsen-chuang}, a natural 
choice of splitting subspaces is provided by 
the one-dimensional joint eigenspaces of the associated stabilizer operators. 
However, as showed in the above Lemma, there is no need to restrict to this 
subset of pure states. 

We also remark that in control-theoretic terms, requirement (s1) corresponds to 
complete controllability of the $N$-qubit in finite time at the level of its Hamiltonian 
description \cite{altafini-tutorial}. In the context of quantum computation, this is 
equivalent to requiring access to a (continuous) set of gates achieving exact 
universality \cite{nielsen-chuang}.  In practice, if unitary operations are built 
out of a \emph{discrete} set of universal gates, an arbitrary target gate $U$ may be 
implemented by a finite-length quantum circuit only within a non-zero accuracy. In 
this case, the finite accuracy of the unitary control action will carry over to the 
stabilized state, with a worst-case error growing linearly in the number of steps.

\vspace*{1mm}

\subsection{Stabilization protocols}

Building on the above result, the next step is to show that the state of the target 
qubits can be prepared in any splitting subspace ${\cal S}$ in one iteration given the 
control resources specified in (s1)-(s3). If we could perform a projective measurement 
of $\Pi_{\cal S},\Pi_{\cal S}^\perp,$ the orthogonal projection onto ${\cal S}$ and its orthogonal complement respectively, it would suffice to use a feedback law that does 
nothing if $\Pi_{\cal S}$ is measured, whereas for the other outcome it applies a 
$U_\perp$ obeying 
\begin{equation}
U_\perp \Pi_{\cal S}^\perp U_\perp^\dag=\Pi_{\cal S} .
\label{perp}
\end{equation}
Notice, again, that such a control action would be highly not unique, since there is
at least enough freedom as associated to the unitary mapping from $\Si^\perp$ to $\Si,$ 
which is an element of $SU(2^{N-1})$.  The average total evolution would then be:
\beq\rho(t+1) \equiv {\cal E}_{{\cal S}}[\rho(t)]
=\Pi_{\cal S}\rho(t)\Pi_{\cal S} + U_\perp\Pi_{\cal S}^\perp\rho(t)
\Pi_{\cal S}^\perp U_\perp^\dag,
\label{Epump}\eeq
and it is easy to see that the support of $\rho(t+1)$ is contained in $\Si$, as 
desired:
\beqan
\trace[\Pi_{\cal S}\rho(t+1) ] &\hspace*{-1.5mm}=\hspace*{-1.5mm}&
\trace[\Pi_{\cal S}\rho(t)] +
\trace[U_\perp^\dag\Pi_{\cal S} U_\perp\Pi_{\cal S}^\perp\rho(t)
\Pi_{\cal S}^\perp)\\
&\hspace*{-1.5mm}=\hspace*{-1.5mm}&
\trace[\Pi_{\cal S}\rho(t)] +\trace[\Pi^\perp_{\cal S}\rho(t)] =1,
\eeqan
where we have taken advantage of Eq. (\ref{perp}). 

Fortunately, the TPCP map ${\cal E}_{\cal S}$ in Eq. (\ref{Epump}) can also be 
implemented without requiring projective measurements, by using the extra qubit as 
a quantum controller. 
Let $\{\ket{\psi_k}\}$ denote an orthonormal basis for $\Hi_Q$, such that the first 
$2^{N-1}$ elements are a basis for $\Si.$  Then proceed as follows:
\begin{itemize}
\item[(1)] Perform a unitary $U_\psi^\dag $ on the target qubits, where 
$U_\psi\ket{\phi_k}=\ket{\psi_k}$.  In this way, the state of (in particular) 
the first qubit contains information on whether or not the state of the system 
is in $\Si$.
\item[(2)] Perform $\CNOT_{\text{out}}$ on the control and the first qubits, thereby 
mapping the information of the first qubit on the state of the control qubit;
\item[(3)] Perform $\CNOT_{\text{in}}$ on the control and the first qubits, thereby 
changing the state of the first qubit depending on the projection of the control 
qubit on $\ket{0}$;
\item[(4)] Perform the change of basis $U_\psi$ to return to the original basis and 
(if needed for successive steps) reset the control qubit back to $|1\rangle$.   
\end{itemize}

\noindent The net effect of operations (1)--(4) is the following unitary transformation 
on $\Hi_c \otimes \Hi_Q$: 
\beqan 
\hspace{-5mm}U_{\text{tot}}&\hspace*{-2.5mm}=\hspace*{-1.5mm}&
U_\psi\CNOT_{\text{in}}\CNOT_{\text{out}}U_\psi^\dag\\
&\hspace*{-2.5mm}=\hspace*{-1.5mm}& 
U_\psi\CNOT_{\text{in}}U_\psi^\dag U_\psi \CNOT_{\text{out}}
U_\psi^\dag\\ 
&\hspace*{-5mm}=\hspace{-5mm}&(\ket{1}\bra{1}\otimes I_{2^{N}}+
\ket{0}\bra{0}\otimes U_\perp)(I_2\otimes\Pi_{\cal S}+\sigma_x\otimes\Pi_{\cal S}^\perp),
\eeqan
where again $U_\perp$ satisfies Eq. (\ref{perp}). Thus, by tracing over the control 
degrees of freedom and assuming that above steps (1)--(4) can be realized in one time unit, this yields:
\begin{eqnarray*}
\rho(t+1)&=&\trace_{\Hi_c} \hspace*{-1mm}\left\{ U_{\text{tot}}
[\ket{1}\bra{1}\otimes\rho(t) ] U_{\text{tot}}^\dag \right\} \\
&=&\Pi_{\cal S}\rho(t)\Pi_{\cal S} + U_\perp\Pi_{\cal S}^\perp\rho(t)
\Pi_{\cal S}^\perp U_\perp^\dag= {\cal E}_{{\cal S}}(\rho_t), 
\end{eqnarray*}
as claimed. The whole sequence of the 4 operations (1)--(4) above will be referred in the following as a {\em control step}, associated to a coherent implementation of a single feedback loop in discrete-time.
If we have $N$ splitting subspaces constructed as in the proof of Lemma \ref{splitsubs}, 
the basis $\{\ket{\psi_k}\}$ to be used at the beginning of the above protocol can be 
chosen {\em at each control step} to be a reordering of the first one.  
With this choice, the desired state-preparation result can then be easily established.

\begin{theorem} \label{thm}
Any $N$-qubit pure state $\rho_d=\ket{\psi}\bra{\psi}$ can be 
deterministically prepared using control resources (s1)--(s3) in $N$ 
steps, irrespective of the initial state $\rho$.  
\end{theorem}

{\em Proof.} Let $\{{\cal S}_\ell\}$ be a splitting-subspace description for 
$\ket{\psi}$ as in Eq. (\ref{SS}), 
with ${\cal E}_{{\cal S}_\ell}$ the associated TPCP map given in \eqref{Epump}.
Now implement the map ${\cal E}_{{\cal S}_N}\circ\ldots\circ{\cal E}_{{\cal S}_1}(\rho).$
By construction, each ${\cal E}_{{\cal S}_\ell}$ maps 
$\left(\bigcap_{k=1}^{\ell-1}{\cal S}_k\right)\cap {\cal S}^\perp_\ell$ 
onto $\left(\bigcap_{k=1}^{\ell-1}{\cal S}_k\right)\cap {\cal S}_\ell$. 
$\left(\bigcap_{k=1}^{N}{\cal S}_k\right)=\textrm{span}\{\ket{\psi}\}$.
Thus, at the $N$-th control step, any initial state is driven into 
$\left(\bigcap_{k=1}^{N}{\cal S}_k\right)=\textrm{span}\{\ket{\psi}\}$.
\qed
We stress that, while the controlled-not operations entangling the control and the 
first qubits, as well as the resetting of the control qubit, are the same at each 
step (thus in real-world implementations will each take a fixed time to be enacted),
{\em the $U_\psi$ will in general differ} at different steps, since $U_\psi$ will 
include a reshuffling of the basis elements in order to obtain the ``nested'' subspace
structure used for the proof.  Since, in general, one can see that up to $(N-\ell)/2$ 
qubit swaps are needed at step $\ell$, this can lengthen the total implementation time, 
making the initial choice of the (non-unique) basis $\{\ket{\psi_k}\}$ important 
for efficient implementation.

\vspace*{1mm}

\subsection{Effective entangling operations}

In practice, engineering exactly the controlled-not gates $ C_{\text{in}}$ and  $C_{\text{out}}$ 
we have assumed may be hard to achieve. Luckily, as it turns out, the way of realizing the ``feedback'' 
map \eqref{Epump} with quantum controllers is highly not unique. Even if the controlled-not 
gates are achievable in principle, different choices may be more convenient to implement, 
motivating us to further characterize the degree of flexibility intrinsic to our protocol.
More precisely, it is easy to verify by direct calculation that {\em any pair} of 
entangling gates of the following form can achieve the desired task up to some additional 
unitary operation on the target qubits: 
\begin{eqnarray}
&\hspace*{-7.5mm}\tilde C_{\text{in}}& \hspace*{-5mm}=
(I_2\otimes  \tilde V_1)\left(\ket{1}\bra{1}\otimes I_2
+\ket{0}\bra{0}\otimes U_\perp\right)  W^\dag (I_2\otimes  \tilde V_2), 
\nonumber \\
&\hspace*{-6mm}\tilde C_{\text{out}}&\hspace*{-3mm} =
(I_2\otimes \tilde U_1) W\left( D_c\otimes \Pi_{0}+
O_c\otimes \Pi_{1}^\perp\right)(I_2\otimes \tilde U_2),
\label{cout}
\end{eqnarray}
where $D_c$, ($O_c$) are unitary and diagonal (off-diagonal) in the standard basis 
of $\Hi_c,$  $\tilde U_j$ and $\tilde V_j$ are unitary operators on $\Hi_Q$  and $W$ is an 
arbitrary unitary operator on the whole system, respectively. $\Pi_0$ $(\Pi_1)$ indicates the 
orthogonal projections on the system subspaces corresponding to the first qubit being 0 (1). 
By invoking assumption (s1), $\tilde U_j$ and $\tilde V_j$ can be compensated by suitable 
unitary control: if, in the steps (2) and (3) of the protocol described above, we now apply 
$(I_2\otimes \tilde U_{1}^\dag)\tilde C_{\text{out}}(I_2\otimes \tilde U_2^\dag)$ and 
$(I_2\otimes \tilde V_1^\dag )\tilde C_{\text{in}}(I_2\otimes \tilde V_2^\dag ),$ respectively, 
it follows that 
\beqan&&\hspace*{-1.3cm}\tilde U_{\text{tot}}\hspace*{-0.5mm}=\hspace*{-.5mm}
U_\psi (I_2\otimes \tilde V_1^\dag) \tilde\CNOT_{\text{in}} (I_2\otimes \tilde V_2^\dag) 
(I_2\otimes \tilde U_1^\dag)\tilde\CNOT_{\text{out}}(I_2\otimes \tilde U_2^\dag) U_\psi^\dag\\
&&\hspace*{-7mm}= \left(\ket{1}\bra{1}\otimes I_{2^N}
+\ket{0}\bra{0}\otimes U_\perp\right)\left( D_c\otimes \Pi_{\cal S}+
O_c\otimes \Pi_{\cal S}^\perp\right).
\eeqan
The net reduced dynamics on the system is not affected by the phases introduced on the ancilla by 
$D_c,\,O_c,$ hence tracing out the ancilla we recover \eqref{Epump} again.

In general, failing to compensate the action of  $\tilde U_j,\tilde V_j$ as described 
could  potentially slow down (or even prevent) the desired state preparation since 
it need no longer be true that 
$\left(\bigcap_{k=1}^{\ell-1}{\cal S}_k\right)\cap {\cal S}^\perp_\ell$ is 
mapped onto $\left(\bigcap_{k=1}^{\ell-1}{\cal S}_k\right)\cap {\cal S}_\ell,$ 
a key step in the convergence proof of Theorem \ref{thm}.  Such compensation 
may, however, be unnecessary if additional conditions are obeyed.  For instance, if 
$\Pi_{{\cal S},k}$ is the projector on the $k$-th splitting subspace ${\cal S}_k,$ and it holds that:
\beqa\tilde V_1,\Pi_{{\cal S},k}&=&0,\label{cc1}\\
\tilde V_1\Pi_{{\cal S},k}&=&\Pi_{{\cal S},k}\label{cc2},\eeqa
then there is no need to compensate for $\tilde V_1.$ In fact, \eqref{cc1} ensures ${\cal S}_k$ 
is still an invariant stable subspace for the current control step, and \eqref{cc2} that $\tilde V$ acts 
as the identity on it, without compromising the result of the previous steps. 
A concrete example will be provided in the following section.
Similar conditions can be worked out for $\tilde U_1,\tilde U_2,\tilde V_2.$

\vspace*{2mm}

\section{Case Study: Trapped Ions}

\subsection{Experimental Bell-state pumping}
\label{exp}

Recently, a toolbox for engineering the open-system dynamics of up to five 
qubits has been experimentally demonstrated using ${}^{40}$Ca ions trapped in 
a linear geometry \cite{barreiro}. The control resources specified in (s1)--(s3)
are available through a combination of single- and multi-qubit gates 
ensuring universal scalable ion-trap quantum computation, plus access to a controllable 
dissipative mechanism on an extra ancilla qubit, realized via a combination of
optical pumping and spontaneous emission.  Here, we revisit the implemented 
two-qubit ``Bell-state pumping'' protocol in the light of our splitting-subspace 
approach. Experimentally, a similar approach has been successfully employed to 
create GHZ states on up to 4 qubits. Let us denote the four Bell states as
\begin{equation}\label{eq:Bell}
|\Phi^\pm\rangle = \frac{1}{\sqrt{2}}(|00\rangle \pm |11\rangle), 
\ |\Psi^\pm\rangle = \frac{1}{\sqrt{2}}(|01\rangle \pm |10\rangle).
\end{equation}
The system, initially in an unknown state specified by a density operator  
$\rho$, is deterministically prepared in the Bell state $|\Psi^-\rangle$ by 
realizing in two steps a quantum operation $\rho \mapsto |\Psi^-\rangle\langle \Psi^-|$. 
It is well known that Bell states are simple examples of stabilizer 
states\footnote{Following standard notation \cite{nielsen-chuang}, we shall 
denote by $X_k,Y_k,Z_k$ the multi-qubit operators that act as Pauli matrices 
$\sigma_x,\sigma_y,\sigma_z$ on the $k$-th qubit, and as identity on the rest.}: 
each of the four Bell states in (\ref{eq:Bell}) may be uniquely characterized 
as a joint eigenstate (with eigenvalues $\pm 1$) of a mutually commuting set of 
two stabilizer generators, for instance $Y_1Y_2$ and $X_1X_2$. The considered 
strategy engineers two maps under which the system qubit state is 
transferred from the $+1$ into the $-1$ eigenspace of $Y_1Y_2$ and $X_1X_2$.

In order to implement the first map, three unitary operations and a dissipative 
one have been used.  All act on the two system qubits, along with the ancillary 
control qubit, denoted as before with the subscript $c$.  
A key role is played by so-called M\o lmer-S\o rensen (MS) entangling gates 
\cite{MSgate}, of the form $U_{X^2}(\theta) = \exp(-i\frac{\theta}{4}S_x^2)$ and 
$U_{Y^2}(\theta) =  \exp(-i\frac{\theta}{4}S_y^2)$, where $S_x$ and $S_y$ denote 
collective spin operators, $S_x \equiv \sum_{i} X_i$ and $S_y \equiv \sum_{i} Y_i$.
The mapping steps of the experimental pumping protocols are then as follows: 

\begin{itemize}
\item[(i)] Information about whether the system is in the $+1$ or $-1$ eigenspace 
of $X_1 X_2$ is mapped by the MS gate $U_{X^2}(\pi/2)$ on $\Hi_c \otimes \Hi_Q$ 
onto the logical states $|0\rangle$ and $|1\rangle$ of the ancilla (initially in 
$|1\rangle$).

\item[(ii)] A controlled gate performs a conversion from the $+1$ eigenvalue of the 
stabilizer $X_1 X_2$ to $-1$ by acting on the ancilla and the first system 
qubits\footnote{In the experiment, the following sequence is employed: 
$C_{\text{in}}=C(p)= U_{Z_1}(\alpha) U_Y(\pi/2)U_{X^2}^{(c,1)}(-\alpha)U_Y(-\pi/2)$, 
with $U_{X^2}^{(c,1)}(-\alpha)= \exp(i(\alpha/2)X_c X_1),$ 
$U_{Z_1}(\alpha) = e^{i\alpha Z_1}$, and $p=\sin^2(\alpha).$}: $$
C_{\text{in}} = |0\rangle\langle 0|_c \otimes Z_1 + |1\rangle\langle 1|_c \otimes I.  $$

\item[(iii)] The MS gate $U_{X^2}(\pi/2)$ is re-applied, in order to move the state 
back to the initial basis representation.

\item[(iv)] The ancilla qubit is dissipatively reset to state $|1\rangle$. 
\end{itemize}
\noindent 
Next, the above cycle is repeated, this time using $U_{Y^2}(\pi/2)$ gates  
in steps (i) and (iii), with the controlled-gate 
$C_{\text{in}}$ remaining unchanged.  

\vspace*{1mm}

\subsection{Bell-state pumping and splitting subspaces} 

In order to show that the above stabilization scheme can be seen as an 
instance of our splitting-subspace approach, it is necessary to 
take a closer look at the structure of the MS gate. The MS entangling gate 
is based on pairwise 
interaction terms and for the present discussion it suffices to consider 
the explicit form already given earlier\footnote{In its most general form, the MS gate 
can be parametrized by two angles $\theta$ and $\phi$, $U_{\text{MS}}(\theta, \phi) = 
\exp\left(\hspace*{-0.5mm}-i\frac{\theta}{4}(\cos\phi \,S_x+ \sin\phi\, S_y)^2\right)$.}, 
that is,  
\begin{equation}
U_{X^2}(\theta) = \exp\left(\hspace*{-.5mm}-i\frac{\theta}{4} S_x^2 \right) \equiv
\exp\Big(\hspace*{-.5mm}-i\frac{\theta}{4}\Big(\sum_i X_i\Big)^2\Big), 
\label{MS}
\end{equation}
where the sum defining the collective spin operators $S_x$ 
is understood to be performed over all the ions involved in the gate. 
In the experimental Bell-state pumping protocol we have just examined, 
MS gates with a phase angle $\theta = \pi/2$ are required. In this case, the 
$U_{X^2}(\pi/2)$ operator in (\ref{MS}) can be decomposed in a more explicit 
form with respect to the relevant 
$\Hi_c \otimes \Hi_Q$ tensor-product decomposition:
$$U_{X^2}\Big(\frac{\pi}{2}\Big) =U'_X (I \otimes \Pi_{-1} + 
X \otimes \Pi_{+1}), $$
where $U'_X$ is a unitary $8\times8$ matrix of the form:
\begin{eqnarray*}
U'_X= I_2\otimes  \frac{1}{2} e^{-i\pi/8}\left[\hspace{-1mm} 
\begin{array}{rrrr} 1 & -1 & -1 & -1 \\ 
-1 & 1 & -1 & -1 \\ -1 & -1 & 1 & -1 \\ -1 & -1 & -1 & 1\end{array} 
\hspace*{-1mm} \right] \equiv I_2 \otimes U''_X. 
\end{eqnarray*}
The action of the unitary operator $U''_X$ on the Bell states basis 
(\ref{eq:Bell}) is described as:
\begin{eqnarray*}
U''_X|\Phi^+\rangle &\hspace*{-1.5mm}=\hspace*{-1.5mm}& -\frac{1}{\sqrt{2}}e^{-i\pi/8} |\Psi^+\rangle ,\\
U''_X|\Phi^-\rangle &\hspace*{-1.5mm}=\hspace*{-1.5mm}& \ \ \frac{1}{\sqrt{2}}e^{-i\pi/8} |\Phi^-\rangle , \\
U''_X|\Psi^+\rangle &\hspace*{-1.5mm}=\hspace*{-1.5mm}& -\frac{1}{\sqrt{2}}e^{-i\pi/8} |\Phi^+\rangle , \\
U''_X|\Psi^-\rangle &\hspace*{-1.5mm}=\hspace*{-1.5mm}& \ \ \frac{1}{\sqrt{2}}e^{-i\pi/8} |\Psi^-\rangle .
\end{eqnarray*}
Hence, $U''_X$ is {\em not} harmful for our purpose since it swaps the Bell 
states in the $+1$ eigenspace of $X_1X_2$ and does not change (except for an 
irrelevant constant), the other Bell states in the $-1$ eigenspace. Explicitly, 
in the Bell basis
$\mathcal{B}_{\text{Bell}} := \{|\Phi^+\rangle, |\Psi^+\rangle, |\Phi^-\rangle, 
|\Psi^-\rangle \}$, $U''_X$ can be written as:
$$ U''_{X,\text{Bell}}=\frac{1}{\sqrt{2}}e^{-i\pi/8} \left[\begin{array}{c|c} 
-X   & O_2 \\ 
\hline O_2 & I_2 \end{array}\right], $$
where the symbol $O_2$ denotes a $2\times2$ matrix of zeros.
We have thus obtained a decomposition of the MS gate $U_{X^2}(\pi/2)$ in a form 
that includes two terms:
\begin{enumerate}
\item The conditional gate $C^X_{\text{out}} = 
I \otimes \Pi_{-1} + X \otimes \Pi_{+1}$, 
that coherently transfers to the ancilla qubit the information on which of the 
two subspaces the system's state is in;
\item An additional unitary $U'_X$ that is not harmful for stabilization purposes, 
since it commutes with the projector onto the eigenspaces of $X_1 X_2$. 
\end{enumerate}
This means that the $U_{X^2}(\pi/2)$ gate can be decomposed precisely in the form 
for $\tilde{C}_{\text{out}}$ given in Eq. \eqref{cout}, and is thus a viable 
operation for implementing a splitting-subspace approach to stabilize 
the desired subspace.

\begin{figure}[t]
\centering
\input{blockdiagram}
\includegraphics[width=8cm]{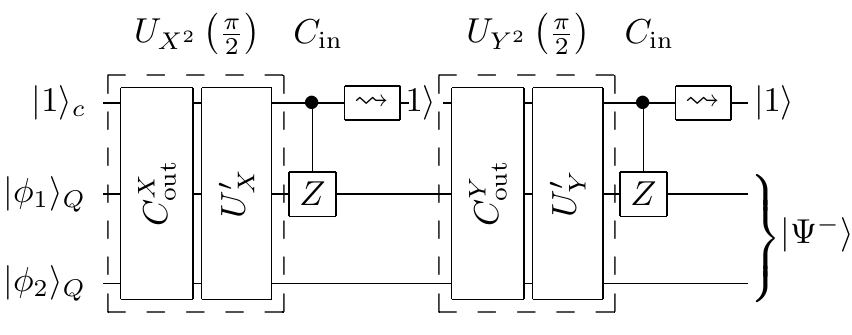} 
\caption{Simplified version (with a {\em single} MS gate being employed in each 
control cycle) of the Bell-state pumping protocol experimentally implemented in 
trapped ions by Barreiro \emph{et al.} \cite{barreiro}.} 
\label{fig:QCirc}
\end{figure}
\vspace*{1mm}

Similarly, the MS gate $U_{Y^2}(\pi/2)$ can be decomposed in the form:
$$ U_{Y^2}\Big(\frac{\pi}{2}\Big) =U'_Y (Z \otimes \Pi'_{-1} + X \otimes \Pi'_{+1}), $$
where now the unitary operator $U'_Y$ reads: 
$$  U'_Y= Z\otimes  \frac{1}{2} e^{-i\pi/8}\left[ \hspace*{-1mm}
\begin{array}{rrrr} 1 & 1 & 1 & \ \ 1 \\ 
-1 & 1 & -1 & 1 \\ -1 & -1 & 1 & 1 \\ 1 & -1 & -1 & 1\end{array} 
\hspace*{-1mm}\right] \equiv Z \otimes U''_Y. $$
Thus, we have likewise found a decomposition of the MS gate $U_{Y^2}(\pi/2)$ in the 
form of Eq. \eqref{cout}, involving a conditional unitary $C^Y_{\text{out}} = 
Z \otimes \Pi'_{-1} + X \otimes \Pi'_{+1}$ and an additional unitary $U'_Y$ with the 
same roles as for the $X_1X_2$ case.  This means that also $U_{Y^2}(\pi/2)$ is 
an admissible entangling operation for the splitting-subspace method.

Note that in the implementation of the experimental protocol described in Sec. \ref{exp}, 
the {\em second} application of the MS gates in step (iii) has essentially the effect of 
canceling the action of $U'_X,U'_Y$ on the target qubits.  However, it is easy to see 
that for the first map, there is no actual need for this second MS gate. In fact, the net 
effect of the ``residual'' $U'_X$  is to swap the states that were in the $+1$ eigenspace 
of $X_1X_2,$ but they end up being pumped in the correct subspace anyway.  Similarly, 
one can check by direct calculation that the action of $U'_Y$ on the subspace that is 
prepared by the first map, namely the one generated by 
$\{|\Phi^-\rangle , |\Psi^-\rangle \}$, is:
\beqan U''_Y|\Phi^-\rangle &\hspace*{-1.5mm}=\hspace*{-1.5mm}& 
-\frac{1}{\sqrt{2}}e^{-i\pi/8} |\Psi^+\rangle \nonumber,\\
U''_Y|\Psi^-\rangle &\hspace*{-1.5mm}=\hspace*{-1.5mm}& \ \ 
\frac{1}{\sqrt{2}}e^{-i\pi/8}|\Psi^-\rangle.\eeqan
The desired state $|\Psi^-\rangle $ is not perturbed: accordingly, if we now apply 
the controlled-$\sigma_z$ to 
the other state we obtain: 
\[ Z_1|\Psi^+\rangle=|\Psi^-\rangle,\]
and the desired state preparation is achieved without the need for step (iii) or, 
equivalently, without undoing $U'_X,U'_Y$.  This simplified version of the 
stabilization protocol, which in fact corresponds to the experimentally implemented 
one as noted in the Supplementary Material \cite{barreiro}, is depicted in Fig. 
\ref{fig:QCirc}. 

\vspace*{2mm}

\section{Conclusions}

We presented a general framework to design stabilizing controls that prepare 
{\em arbitrary pure states on qubit registers in finite time}: once a representation 
of the target state in terms of splitting subspaces is chosen, the control objective 
can be achieved either by measurements and discrete-time {\em classical feedback}
\cite{bolognani-arxiv,bolognani-cyprus}, 
or by resorting to {\em coherent feedback} via quantum controllers 
\cite{lloyd-coherent}.
In the latter case, the measurement and feedback steps are effectively replaced by 
suitable conditional operations ($C_{\text{out}},C_{\text{in}}$ in our setting), which  
enable the required quantum-information flows out of the system to the 
controller, and viceversa.  This fully coherent implementation can be practically 
advantageous in a number of situations where quantum measurements are exceedingly slow 
and/or inaccurate, or destructive for the system itself.  

The proposed splitting-subspace approach leaves, in its current form, significant freedom 
in constructing a suitable basis for the system state space.  While we have shown how 
the latter translates in added flexibility for implementing the required conditional 
operations and can be exploited in principle to minimize the complexity of the 
stabilization protocol, systematic techniques for {\em optimizing} the generation of 
a splitting-subspace decomposition in specific control settings remain an interesting 
problem for future studies.
We have illustrated our general approach by revisiting a recently proposed protocol 
for dissipative Bell-state preparation in ion traps \cite{barreiro}  
in the light of an explicit splitting-subspace analysis, demonstrating its direct 
applicability to current experimental situations.  We expect that this will 
pave the way to further applications of our ideas to synthesizing finite-time 
dissipative state-preparation protocols in other qubit devices and control 
scenarios of experimental relevance.  

\vspace*{2mm}

\section*{Acknowledgements}
L.V. gratefully acknowledges hospitality by the Department of Information Engineering 
of the University of Padua, where part of this work was performed. F.T. wishes to thank 
M. M\"uller and T. Monz for stimulating discussions.

\vspace*{2mm}

\bibliography{bib-francesco-2}
\bibliographystyle{plain}

\end{document}